\documentclass[eqsecnum,prd,aps,preprint,nofootinbib]{revtex4}  
   
\usepackage{graphicx} 
\usepackage{mathrsfs}
\usepackage{amsmath,amsfonts,amssymb}
\usepackage{amsthm}
\theoremstyle{plain} 
\newtheorem{thm}{Theorem}

\theoremstyle{definition}

\theoremstyle{remark}

\begin{document}

\title{From parabolic to loxodromic BMS transformations}

\author{Giampiero \surname{Esposito}$^{1}$}
\author{Francesco \surname{Alessio}$^{1,2}$}

\affiliation{
$^1$INFN Sezione di Napoli, Complesso Universitario di Monte S. Angelo,\\ 
Via Cintia Edificio 6, 80126 Naples, Italy\\
$^2$Dipartimento di Fisica ``Ettore Pancini'', Federico II University, 
Complesso Universitario di Monte S. Angelo,\\
Via Cintia Edificio 6, 80126 Naples, Italy}
 
\date{\today}

\begin{abstract}
Half of the Bondi-Metzner-Sachs (BMS) transformations consist of orientation-preserving
conformal homeomorphisms of the extended complex plane known as fractional linear 
(or M\"{o}bius) transformations. These can be of $4$ kinds, i.e. they are
classified as being parabolic, or hyperbolic, or elliptic, or loxodromic, depending on the
number of fixed points and on the value of the trace of the associated $2 \times 2$ matrix
in the projective version of the $SL(2,\mathbb {C})$ group. The resulting particular forms of
$SL(2,\mathbb {C})$ matrices affect also the other half of BMS transformations, and are used here
to propose $4$ realizations of the asymptotic symmetry group that we call, again,
parabolic, or hyperbolic, or elliptic, or loxodromic. 

In the second part of the paper, we prove that a subset of hyperbolic 
and loxodromic transformations, those having trace that approaches $\infty$, correspond to the 
fulfillment of limit-point condition for singular Sturm-Liouville problems. Thus,  
a profound link may exist between the language for describing asymptotically flat space-times
and the world of complex analysis and self-adjoint problems in ordinary quantum mechanics.
\end{abstract}

\maketitle

\section{Introduction}
\setcounter{equation}{0}

The asymptotic symmetry group of asymptotically flat space-times, originally discovered thanks
to the work \cite{BMS1,BMS2,BMS3} 
of Bondi, Metzner and Sachs (hereafter BMS), is still part of modern research in
gravitational physics, thanks to the theoretical investigation of black hole physics 
\cite{BH1,BH2,BH3,BH4}, Hamiltonian methods \cite{L1,L2,H1,H2,H3,H4} and symmetry groups 
of various space-time models \cite{MC1,MC2,MC3,MC4,MC5,M1,M2,M3,Calo,OB1,CEK1,A1}.

In particular, when a generic metric tensor is expressed in Bondi-Sachs coordinates
$(u,r,x^{B})$, it reads as \cite{A1}
\begin{eqnarray}
g&=& e^{2 \beta} {V \over r}du \otimes du+e^{2 \beta}(du \otimes dr+ dr \otimes du)
\nonumber \\
&+& \sum_{B,C=1}^{2}g_{BC}(dx^{B}-U^{B} \; du) \otimes (dx^{C}-U^{C} \; du).
\label{(1.1)}
\end{eqnarray}
With this notation, $x^{B}$ consists of the pair of coordinates ($\theta,\phi$), and the
BMS transformations are the diffeomorphisms which do not affect the asymptotic form 
of the metric of an asymptotically flat space-time. At a deeper level, the group
of conformal transformations of future null infinity which preserve both angles and
null angles \cite{PR1,PR2,A1} is the BMS group. 
Upon defining the complex stereographic coordinate
\begin{equation}
\zeta \equiv e^{i \phi} \; \cot {\theta \over 2},
\label{(1.2)}
\end{equation}
and considering the $SL(2,\mathbb{C})$ matrix
\begin{equation}
M_{2}=\left(\begin{matrix}
a & b \\ c & d 
\end{matrix}\right), \;
ad-bc=1,
\label{(1.3)}
\end{equation}
the general form of a BMS transformation is well known to be
(${\overline \zeta}$ being the complex conjugate of $\zeta$)
\begin{equation}
\zeta \rightarrow {{a \zeta + b}\over {c \zeta +d}},
\label{(1.4)}
\end{equation}
\begin{equation}
u \rightarrow {(1+|\zeta|^{2}) [u+A(\zeta,{\overline \zeta})] \over
{|a \zeta+b|^{2}+ |c \zeta+d|^{2}}}.
\label{(1.5)}
\end{equation}

Now the simple but non-trivial point of our paper is that the maps \eqref{(1.4)} are the 
fractional linear transformations whose properties are well known in complex analysis and group
theory, but have not been fully exploited by relativists and theoretical physicists. 
Section $2$ outlines the well-established classification of maps \eqref{(1.4)}, 
while in Section $3$ it is shown that an important 
correspondence exists between this classification and 
the usual translations, rotations and boosts that allows to see the whole of the Poincar\'e 
group from a new perspective. Section $4$ exploits this classification to obtain $4$ basic 
forms of the maps \eqref{(1.5)}, that we
call parabolic, hyperbolic, elliptic, loxodromic. Section $5$ obtains an intriguing
link between fractional linear transformations and the geometry of singular
Sturm-Liouville problems, while concluding remarks are presented in Section $6$,
and technical details are provided in the appendix. 

\section{Fractional linear transformations}
\setcounter{equation}{0}

In complex analysis \cite{Simon}, a fractional linear transformation is an orientation-preserving
conformal homeomorphism $h$ of the extended complex plane
$\mathbb{C} \cup \{ \infty \}$, such that
\begin{equation}
h(z)={{{\tilde \alpha}z+{\tilde \beta}}\over
{{\tilde \gamma}z+{\tilde \delta}}}
={{t {\tilde \alpha}z+t {\tilde \beta}}\over
{t {\tilde \gamma}z+t {\tilde \delta}}}, \;
\forall t \not =0.
\label{(2.1)}
\end{equation}
Since the ratio in \eqref{(2.1)} is independent of $t$, one can exploit this to make sure that,
eventually, the matrix of coefficients has unit determinant (we define
$\alpha \equiv t {\tilde \alpha},...,\delta \equiv t {\tilde \delta}$), i.e.
$$
t^{2}\Bigr({\tilde \alpha} \; {\tilde \delta}
-{\tilde \beta} \; {\tilde \gamma}\Bigr)=1 \Longrightarrow 
\left(\begin{matrix}
\alpha & \beta \\
\gamma & \delta
\end{matrix}\right) \in \; {\rm SL}(2,\mathbb{C}),
$$
with associated
\begin{equation}
h(z)={{\alpha z + \beta}\over {\gamma z + \delta}}.
\label{(2.2)}
\end{equation}
The fixed points of $h$ solve, by definition, the equation $h(z)=z$, 
i.e., from \eqref{(2.2)}
\begin{equation}
\gamma z^{2}+(\delta-\alpha)z-\beta=0,
\label{(2.3)}
\end{equation}
which is solved by
\begin{equation}
z={(\alpha-\delta) \pm \sqrt{(\alpha+\delta)^{2}-4(\alpha \delta - \beta \gamma)} \over 2}
={(\alpha-\delta) \pm \sqrt{(\alpha+\delta)^{2}-4} \over 2},
\label{(2.4)}
\end{equation}
where we have exploited the $SL(2,\mathbb{C})$ condition.

Thus, if
\begin{equation}
(\alpha+\delta)^{2}=4 \Longrightarrow |\alpha+\delta|=2,
\label{(2.5)}
\end{equation}
there exists only one fixed point $z={(\alpha-\delta)\over 2}$, and the map $h$ is said 
to be {\it parabolic}. Now a theorem guarantees that every parabolic transformation
can be mapped into a transformation whose only fixed point is at $\infty$, i.e.
\cite{Simon,Bianchi,Maskit}
\begin{equation}
h_{P}(z)=z+\beta.
\label{(2.6)}
\end{equation}
This is a translation in $\mathbb{C} \cup \{ \infty \}$, and is not periodical. The
representative matrix in $PSL(2,\mathbb{C})$ is
\begin{equation}
M_{P}=\left(\begin{matrix} 
\pm 1 & \beta \\
0 & \pm 1
\end{matrix}\right).
\label{(2.7)}
\end{equation}

If the discriminant $(\alpha+\delta)^{2}-4$ in \eqref{(2.4)} does not vanish, two fixed points
are instead found to occur, and the resulting homeomorphism $h$ can be mapped into a
transformation with fixed points at $0$ and $\infty$, i.e. \cite{Bianchi,Maskit}
\begin{equation}
h(z)={\alpha \over \delta}z=kz, \; \alpha \delta=1.
\label{(2.8)}
\end{equation}
Hence one finds
\begin{equation}
{\alpha \over \delta}=\alpha^{2}=k \Longrightarrow \alpha=\sqrt{k}
\Longrightarrow j \equiv {\rm tr}(h)=\alpha+\delta=\alpha+{1 \over \alpha}
=\sqrt{k}+{1 \over \sqrt{k}}.
\label{(2.9)}
\end{equation}
Now one can distinguish three cases:
\vskip 0.3cm
\noindent
(i) If $k=|\kappa|>0$, $h$ is said to be {\it hyperbolic}, and for it
\begin{eqnarray}
\; & \; & h(z)=h_{H}(z)=|\kappa|z, 
\nonumber \\
& \; & {\rm tr}^{2}(h)-4=(\alpha+\delta)^{2}-4 >0 \Longrightarrow |\alpha+\delta| > 2,
\label{(2.10)}
\end{eqnarray}
and the corresponding $2 \times 2$ matrix is
\begin{equation}
M_{H}=\left(\begin{matrix}
|\kappa| & 0 \\
0 & 1
\end{matrix}\right).
\label{(2.11)}
\end{equation}
Equation \eqref{(2.10)} for $h_{H}(z)$  
is a dilation of the plane, and under its action all lines passing
through the origin remain fixed. 
\vskip 0.3cm
\noindent
(ii) If $k \in \mathbb{C}$ and $|k|=1$, one can write 
\begin{equation}
k=e^{i \varphi} \Longrightarrow j=e^{i{\varphi \over 2}}
+e^{-i{\varphi \over 2}}=2 \cos {\varphi \over 2}.
\label{(2.12)}
\end{equation}
The transformation $h$ is then said to be {\it elliptic}, and for it
\begin{equation}
{\rm tr}^{2}(h)=(\alpha+\delta)^{2}<4 \Longrightarrow |\alpha+\delta| < 2,
\label{(2.13)}
\end{equation}
\begin{equation}
h(z)=h_{E}(z)=e^{i \varphi} \; z,
\label{(2.14)}
\end{equation}
with resulting matrix 
\begin{equation}
M_{E}=\left(\begin{matrix}
e^{i \varphi} & 0 \\
0 & 1
\end{matrix}\right).
\label{(2.15)}
\end{equation}
The map \eqref{(2.14)} can be {\it periodic} provided that there exists a natural number $l$ such
that $\varphi=2 \pi l$. A normal elliptic \cite{Bianchi} transformation \eqref{(2.14)}
is a rotation of the complex plane about the origin, with amplitude $\varphi$,
and for it all circles centred at the origin remain fixed.
\vskip 0.3cm
\noindent
(iii) If $k=\rho \; e^{i \sigma} \in \mathbb{C}$, one finds from \eqref{(2.9)}
\begin{equation}
j=\left(\sqrt{\rho}+{1 \over \sqrt{\rho}}\right) \cos {\sigma \over 2}
+i \left(\sqrt{\rho}-{1 \over \sqrt{\rho}}\right) \sin {\sigma \over 2} 
\; \; \in \mathbb{C},
\label{(2.16)}
\end{equation}
while if $k$ is $<0$ one finds, again from \eqref{(2.9)},
\begin{equation}
j={1-|k| \over i \sqrt{|k|}} \; \; \in \mathbb{C}.
\label{(2.17)}
\end{equation}
The resulting transformation $h_{L}$ is said to be {\it loxodromic}, and it
reads as
\begin{equation}
h(z)=h_{L}(z)=\rho \; e^{i \sigma} \; z,
\label{(2.18)}
\end{equation}
with corresponding $2 \times 2$ matrix 
\begin{equation}
M_{L}=\left(\begin{matrix}
\rho e^{i \sigma} & 0 \\
0 & 1
\end{matrix}\right).
\label{(2.19)}
\end{equation}
A loxodromic transformation is obtained by combining an elliptic and a hyperbolic 
transformation, with the same fixed points \cite{Bianchi}. Some authors say that a
non-elliptic transformation with exactly two fixed points is loxodromic, and that 
these include the hyperbolic transformations \cite{Maskit}.

In figure $1$ we describe the various families of fractional linear transformations 
in the complex-$j^{2}$ plane.

\begin{figure}
\includegraphics[scale=0.90]{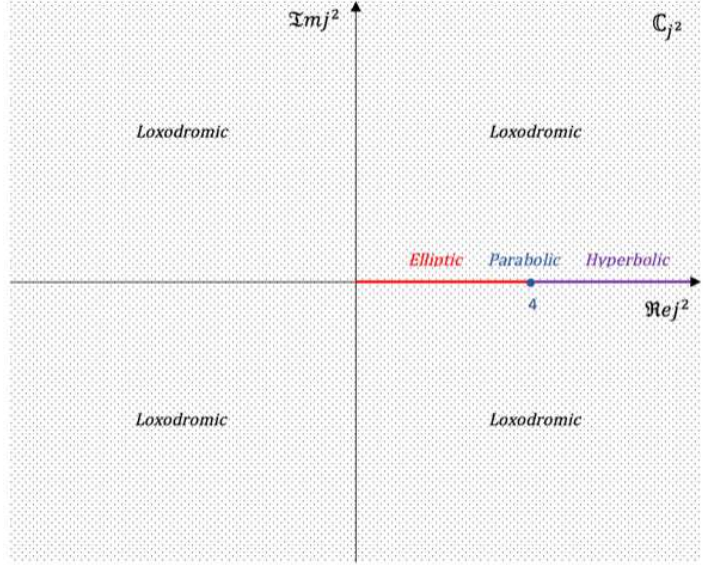}
\caption{If the squared trace $j^{2}$ is used for the classification, fractional
linear transformation are elliptic if $j^{2}<4$, parabolic if $j^{2}=4$, hyperbolic
if $j^{2}>4$, and loxodromic if $j^{2}$ is not real-valued \cite{Simon}.}
\end{figure}

\section{Connection with the Poincar\'e group}

The aim of this Section is to show that the classification of fractional linear 
transformations carried out in Sec. II,
setting momentarily aside the BMS group,
has a deep relation with special relativity. 
In particular, we would like to show that the parabolic, hyperbolic and elliptic 
transformations are strictly related to the usual concepts of translations, boosts and 
rotations respectively, which are the building blocks of the Poincar\'e group. Hence, 
before proceeding to the classification of the BMS group that, as mentioned, is the 
asymptotic symmetry group of asymptotically flat space-times, we regard it interesting 
to see explicitly such a connection with flat space-time symmetry group. Let us start with translations. \\
Consider two complex variables, $z$ and $w$, related to Minkowski space-time coordinates $x^{\mu}$ by 
\begin{align*}
z=x^0+ix^1,\\ 
w=x^2+ix^3.
\end{align*}
We stress that $z$ and $w$ are \textit{not} interpreted here as stereographic coordinates 
on the celestial sphere. Consider the action on them by means of two parabolic transformations, 
$h$ and $h'$, of parameters $\beta$ and $\delta$ respectively:
\begin{subequations}
\begin{align}
\label{(3.1a)}
&z'=h_P(z)=z+\beta, \\
\label{(3.1b)}
&w'=h'_P(w)=w+\delta.
\end{align}
\end{subequations}
Then, if we make the identifications
\begin{align*}
\beta=a^0+ia^1,\\
\delta=a^2+ia^3,
\end{align*}
then eqs. \eqref{(3.1a)} and \eqref{(3.1b)} obviously induce the Minkowski four-translation
\begin{equation}
\label{(3.2)}
x'^{\mu}=x^{\mu}+a^{\mu}.
\end{equation}
Hence, as we claimed, 
\textit{there exists an isomorphism between two copies of the subgroup of parabolic 
transformations and the $4$-translations}.\\
As is clear, this result relies on the trivial isomorphism $\mathbb{C}\simeq\mathbb{R}^2$.\\
Now, turn attention to rotations and boosts. It is well known that, in virtue of the isomorphism 
$L^{\uparrow}_+\simeq SL(2,\mathbb{C})/\mathbb{Z}_2$, where $L^{\uparrow}_+$ denotes the 
connected component of $SO(1,3)$, a rotation of an angle $\phi$ and a boost of rapidity 
$\chi$ about an axis $\hat{n}=(\cos\varphi\sin\theta,\sin\varphi\sin\theta,\cos\theta)$ 
can be performed by using $SL(2,\mathbb{C})$ matrices
\begin{subequations}
\begin{align}
\label{(3.3a)}
&U_{\hat{n}}(\phi)=e^{\frac{i}{2}\vec{\sigma}\cdot\hat{n}\phi}=\mathbb{I}\cos(\phi/2)
+i(\vec{\sigma}\cdot\hat{n})\sin(\phi/2),\\
\label{(3.3b)}
&H_{\hat{n}}(\chi)=e^{\frac{1}{2}\vec{\sigma}\cdot\hat{n}\chi}=\mathbb{I}\cosh(\chi/2)
+(\vec{\sigma}\cdot\hat{n})\sinh(\chi/2).
\end{align}
\end{subequations}
Note that also $-U_{\hat{n}}$ and $-H_{\hat{n}}$ can be used as well.\\
We can take the traces to obtain
\begin{subequations}
\label{(3.4)}
\begin{align}
\label{(3.4a)}
&j_U\equiv\mathrm{tr}(U_{\hat{n}}(\phi))=2\cos(\phi/2)\Longrightarrow j_U^2\leq 4,\\
\label{(3.4b)}
&j_H\equiv\mathrm{tr}(H_{\hat{n}}(\chi))=2\cosh(\chi/2)\Longrightarrow j_H^2\geq 4.
\end{align}
\end{subequations}
Then it follows that \textit{rotations correspond to elliptic transformations} and that 
boosts correspond to \textit{hyperbolic transformations}. The limit cases in which in Eqs. 
\eqref{(3.4)} there holds the equality ($\phi=2k\pi$, where $k\in\mathrm{Z}$, and $\chi=0$) 
correspond again to parabolic transformations.\\
We can eventually state that, from the above analysis, the whole Poincar\'e group 
can be expressed in terms of fractional linear transformations.

\section{From parabolic to loxodromic BMS transformations}
\setcounter{equation}{0}

As is clear from \eqref{(1.4)}, the choice of coefficients $a,b,c,d$ affects also 
the other half of BMS transformations, given by Eq. \eqref{(1.5)}. Thus, bearing in mind 
the matrices \eqref{(2.7)}, \eqref{(2.11)}, \eqref{(2.15)} and \eqref{(2.19)}, our next 
logical step is to write Eq. \eqref{(1.5)} in the form
\begin{equation}
u \rightarrow F(\zeta,{\bar \zeta}) \cdot \Bigr[u+A(\zeta,{\bar \zeta}) \Bigr],
\label{(4.1)}
\end{equation}
where we distinguish $4$ cases as follows.
\vskip 0.3cm
\noindent
(I) Parabolic BMS:
\begin{equation}
\zeta \rightarrow \pm \zeta+\beta,
\label{(4.2)}
\end{equation}
\begin{equation}
u \rightarrow F_{P} \cdot \Bigr[u+A \Bigr], \;
F_{P}={(1+|\zeta|^{2}) \over \Bigr[1+|\pm \zeta+\beta|^{2}\Bigr]}.
\label{(4.3)}
\end{equation}
\vskip 0.3cm
\noindent
(II) Hyperbolic BMS:
\begin{equation}
\zeta \rightarrow |\kappa| \; \zeta,
\label{(4.4)}
\end{equation}
\begin{equation}
u \rightarrow F_{H} \cdot \Bigr[u+A \Bigr], \;
F_{H}={(1+|\zeta|^{2}) \over \Bigr(1+|\kappa|^{2} \; |\zeta|^{2}\Bigr)}.
\label{(4.5)}
\end{equation}
\vskip 0.3cm
\noindent
(III) Elliptic BMS:
\begin{equation}
\zeta \rightarrow e^{i \varphi} \; \zeta,
\label{(4.6)}
\end{equation}
\begin{equation}
u \rightarrow F_{E} \cdot \Bigr[u+A \Bigr], \;
F_{E}={{1+|\zeta|^{2}} \over \Bigr[|\zeta \; e^{i \varphi}|^{2}+1 \Bigr]}=1.
\label{(4.7)}
\end{equation}
Note that, after a {\it periodic} elliptic transformation (see the definition
after (2.15)), Eq. (4.6) implies that the complex variable $\zeta$ defined in (1.2)
remains unaffected, i.e.
$$
\zeta'=e^{2 \pi l i} \zeta=\zeta,
$$
because $\phi$ is mapped into $\phi+2 \pi l$, and the phase is an integer 
multiple of $2 \pi$. In particular, any function $A(\zeta,\bar{\zeta})$ is invariant under 
such a transformation. Thus, once {\it a choice of the function $A$ is made}, Eqs.
(4.6) and (4.7) describe the supertranslation
$$
\theta \rightarrow \theta, \; \phi \rightarrow \phi, \;
u \rightarrow u+A(\zeta,{\bar \zeta}).
$$
\vskip 0.3cm
\noindent
(IV) Loxodromic BMS:
\begin{equation}
\zeta \rightarrow \rho \; e^{i \sigma} \; \zeta,
\label{(4.8)}
\end{equation}
\begin{equation}
u \rightarrow F_{L} \cdot \Bigr[u+A \Bigr], \;
F_{L}={(1+|\zeta|^{2})\over \Bigr(1+\rho^{2}|\zeta|^{2}\Bigr)}.
\label{(4.9)}
\end{equation}
We remark the complete formal analogy between $F_{H}$ and $F_{L}$, in agreement
with the previously stated (but apparently unrelated) property, according to which
loxodromic transformations include the hyperbolic family \cite{Maskit}.  

\section{Fractional linear transformations and singular Sturm-Liouville problems}
\setcounter{equation}{0}

Since our paper is aimed at finding links between well-established but apparently
unrelated branches of mathematics and physics, 
we here consider the fractional linear transformations
that pertain to singular Sturm-Liouville problems, and then try to exploit the
classification studied in Section $2$.

A linear homogeneous second-order differential equation in one real variable $x$ can be
always reduced to the (canonical) form \cite{Whittaker, Esposito}
\begin{equation}
L_{Q} \; u(x)=0, \; \;
L_{Q} \equiv -{d^{2}\over dx^{2}}+Q(x),
\label{(5.1)}
\end{equation}
while a regular Sturm-Liouville problems consists of an eigenvalue equation for the 
linear operator $L_{q} \equiv -{d^{2}\over dx^{2}}+q(x)$, i.e.
\begin{equation}
L_{q} \; u(x)=\lambda \; u(x) \; \; x \in [a,b],
\label{(5.2)}
\end{equation}
where $q$ is a suitably smooth potential term, supplemented by the following boundary
conditions at the ends of the interval (hereafter $\omega \in [0,\pi[$ and 
$\eta \in [0,\pi[$):
\begin{align}
&u(a) \cos(\omega)+u'(a)\sin(\omega)=0,\\
&u(b)\cos(\eta)+u'(b)\sin(\eta)=0.
\end{align}
For the problem (5.2)-(5.4), there exists a discrete spectral resolution of the
one-dimensional Laplace-type operator $L_{q}$, with a countable infinity of real
eigenvalues, and the eigenfunctions forming a complete basis for the space $L^{2}(a,b)$.

The theory of singular Sturm-Liouville problems deals with the case where the closed
interval $[a,b]$ is replaced by the interval $[0,\infty[$, by studying a sequence of
intervals $[a,b[$ as $b \rightarrow \infty$. The pioneering work of Weyl made it possible
to prove the following results \cite{Weyl,Coddington}:
\vskip 0.3cm
\noindent
\begin{thm} $\\ $
\label{thm1}
 If ${\rm Im}(\lambda) \not =0$, and $\chi,\psi$ are the linearly independent
solutions of the eigenvalue equation $L_{q} \; u=\lambda \; u$ which satisfy
\begin{align}
&\chi(0,\lambda)=\sin(\omega),\hspace{1cm}\chi'(0,\lambda)=-\cos(\omega),\\
&\psi(0,\lambda)=\cos(\omega),\hspace{0.9cm}\psi'(0,\lambda)=\sin(\omega),
\end{align}
then there exists a function $m$ of $\lambda$ such that, for all 
$\lambda \in \mathbb{C}- \mathbb{R}$, the linear combination
\begin{equation}
y(x,\lambda)=\chi(x,\lambda)+m(\lambda)\psi(x,\lambda)
\label{(5.7)}
\end{equation}
satisfies the real boundary condition
\begin{equation}
y(b)\cos(\eta)+y'(b)\sin(\eta)=0.
\label{(5.8)}
\end{equation}
\end{thm}
Upon defining
\begin{equation}
A \equiv \chi(b,\lambda), \;
B \equiv \chi'(b,\lambda), \;
C \equiv \psi(b,\lambda), \;
D \equiv \psi'(b,\lambda),
\label{(5.9)}
\end{equation}
\begin{equation}
z \equiv \cot \eta,
\label{(5.10)}
\end{equation}
the boundary condition (5.8) leads to the fractional linear transformation
\begin{equation}
m=-{{Az+B}\over (Cz+D)}, 
\label{(5.11)}
\end{equation}
and hence
\begin{equation}
z=-{{Dm+B}\over (Cm+A)},
\label{(5.12)}
\end{equation}
where $AD-BC=1$ by virtue of Theorem \ref{thm1}, as is proved in Ref. \cite{Coddington}.
The reality condition for the variable $z$ defined in (5.10) is therefore the
equation of a circle $C_{b}$:
\begin{equation}
|m|^{2}-{({\bar B}C-{\bar A}D)\over ({\bar C}D-C{\bar D})}m
-{(A{\bar D}-B {\bar C})\over ({\bar C}D-C{\bar D})}{\bar m}
+{({\bar A}B-A{\bar B})\over ({\bar C}D-C{\bar D})}=0,
\label{(5.13)}
\end{equation}
having centre
\begin{equation}
M_{b}={{A{\bar D}-B{\bar C}}\over ({\bar C}D-C{\bar D})},
\label{(5.14)}
\end{equation}
and radius
\begin{equation}
r_{b}={1 \over |{\bar C}D-C{\bar D}|}.
\label{(5.15)}
\end{equation}
The following theorem holds:
\begin{thm} $\\ $ As $b$ approaches $\infty$, either $C_{b} \rightarrow C_{\infty}$, a
{\it limit circle}, or $C_{b} \rightarrow m_{\infty}$, a {\it limit point}.
In the limit circle case, all solutions of the eigenvalue equation
$L_{q} \; u=\lambda \; u$ are of class $L^{2}(0,\infty)$. If 
${\rm Im}(\lambda)\not =0$, only one solution of $L_{q} \; u=\lambda \; u$ is
of class $L^{2}(0,\infty)$ in the limit-point case.
\end{thm}

At this stage, on the right-hand side of Eq. (5.11) we multiply numerator and
denominator by a non-vanishing complex number $\tau$, and obtain
\begin{equation}
m={{\alpha z + \beta}\over (\gamma z + \delta)},
\label{(5.16)}
\end{equation}
where
\begin{equation}
\alpha \equiv -\tau A, \; \beta \equiv -\tau B, \;
\gamma \equiv \tau C, \;
\delta \equiv \tau D.
\label{(5.17)}
\end{equation}
Thus, in light of the condition $AD-BC=1$, the matrix
on the right-hand side of (5.16) is in $SL(2,\mathbb{C})$ if and only if
$\tau=\pm i$. With the resulting form of the matrix, i.e.
\begin{equation}
\left(\begin{matrix}
\alpha & \beta \\
\gamma & \delta
\end{matrix}\right)
=\left(\begin{matrix}
\mp i A & \mp i B \\
\pm i C & \pm i D 
\end{matrix}\right),
\label{(5.18)}
\end{equation}
we can exploit the classification of section $2$ from which, for the trace 
$j \equiv \alpha + \delta$, we obtain the correspondence \cite{Simon,Bianchi}
\begin{align*}
&j\in \mathbb{R}, \hspace{0.1cm} |j|<2 \hspace{0.1cm} \Longrightarrow \hspace{0.1cm} {\rm elliptic},\\
&j\in \mathbb{R}, \hspace{0.1cm} |j|=2 \hspace{0.1cm} \Longrightarrow \hspace{0.1cm}{\rm parabolic},\\
&j\in \mathbb{R}, \hspace{0.1cm} |j| >2 \hspace{0.1cm} \Longrightarrow \hspace{0.1cm} {\rm hyperbolic},\\
&j\in \mathbb{C}-\mathbb{R} \hspace{0.1cm} \Longrightarrow \hspace{0.1cm} {\rm loxodromic}.
\end{align*}
From Eq. (5.15), the limit-point case, for which the radius $r_{b}$ of the circle 
$C_{b}$ approaches $0$, reduces to
\begin{equation}
|{\bar C}D-C{\bar D}|=|{\bar \gamma} \; \delta-\gamma \; {\bar \delta}| 
\rightarrow \infty,
\label{(5.19)}
\end{equation}
i.e.
\begin{equation}
\left | {\rm Im}(\gamma \; {\bar \delta}) \right | \rightarrow \infty
\; \Longrightarrow \left | {\rm Im} \gamma ({\bar \alpha}-{\bar j})\right |
\rightarrow \infty.
\label{(5.20)}
\end{equation}
This shows that the loxodromic or hyperbolic sectors, which contain also fractional
linear transformations having $|j| \rightarrow \infty$, may lead to shrinkage to zero
of $r_{b}$ and hence to the limit-point condition. Alternatively, one might assume
that $|\gamma| \rightarrow \infty$, but this is incompatible with the desire of having
square-integrable eigenfunctions of the operator $L_{q}$, as is clear from (5.9)
and (5.17).

On the other hand, as far as the limit-circle condition is concerned, all elliptic and 
parabolic fractional linear transformations are acceptable.

Although the established correspondence between singular Sturm-Liouville problems and 
fractional linear transformations is not in the $1-1$ form, it shows (in our 
opinion) an intriguing link between the modern theory of ordinary differential equations
on the one hand, and fractional linear and BMS transformations on the other hand.

The limit-point condition considered in Section $5$ is of particular interest because, for
eigenvalue problems on $(0,\infty)$, the limit-point condition 
at $0$ and $\infty$ is the necessary and sufficient condition (see appendix) 
for proving essential self-adjointness on 
$C_{0}^{\infty}(0,\infty)$ of the radial part of the quantum mechanical Hamiltonian
in a central potential \cite{Reed,Simon4}. In other words, relying upon separately
well-established properties of real, complex and functional analysis on the one hand
and asymptotic structure of general relativity on the other hand, 
we are suggesting that a profound link may exist
between self-adjoint problems in ordinary quantum mechanics and the fractional linear and
hence BMS transformations of general relativity in the large-trace loxodromic and
hyperbolic sectors.

Further applications of projective $SL(2,\mathbb{C})$ transformations have been discovered in
Ref. \cite{Barrella}, where the authors study three-dimensional 
anti-de Sitter space and find that the conformal boundary is acted upon precisely by elements
of $PSL(2,\mathbb{C})$.

\section{Concluding remarks and open problems}
\setcounter{equation}{0}

As far as we can see, a synthesis of our findings is as follows.

The fractional linear transformations (hereafter FLT) of complex analysis are the appropriate
tool for expressing several concepts, i.e.
\vskip 0.3cm
\noindent
(i) The $4$-translations, rotations and boosts of special relativity correspond to $2$ copies
of parabolic FLT, or elliptic FLT, or hyperbolic FLT, respectively.
\vskip 0.3cm
\noindent
(ii) The parabolic through loxodromic FLT engender the bigger group of parabolic through loxodromic
BMS transformations. 
\vskip 0.3cm
\noindent
(iii) The limit-circle condition of singular Sturm-Liouville problems corresponds to
elliptic and parabolic FLT, plus loxodromic and hyperbolic FLT with finite trace $j$.
The limit-point condition of singular Sturm-Liouville problems corresponds instead to loxodromic 
and hyperbolic FLT having trace that approaches $\infty$.

This means that, for example, a Hamiltonian with equal but non-vanishing deficiency indices
\cite{Reed} in ordinary quantum mechanics is 
``dual'' to elliptic and parabolic BMS transformations, 
including BMS supertranslations. On the other side, an essentially self-adjoint 
Hamiltonian is ``dual'' to the large-trace subset of loxodromic and hyperbolic BMS
transformations.

If the mathematical language of asymptotically flat space-times is naturally ``dual''
to the singular Sturm-Liouville problems of ordinary quantum mechanics, the important
question arises of whether the language of full general relativity is
the unexpected gateway to the world of full quantum field theory, and which tools replace FLT 
in the affirmative case.

In a different framework, an example of gateway between general relativity and quantum fields
is provided by the massless Rarita-Schwinger equations. 
When these are studied in curved space-time,
the integrability conditions for finding gauge-invariant solutions of such equations
lead to Ricci flatness, which is equivalent to solving the vacuum Einstein equations
\cite{Penrose} with the exception of two-dimensional space-time.
Thus, maybe a new perspective in theoretical physics might be the task of establishing
correspondences between {\it different} areas of classical and quantum physics, rather
than the attempt of quantizing or dequantizing. We hope that the resulting landscape
awaiting discovery, if it exists, will become accessible in the years to come. Another
related question is whether a rigorous theory of discrete gravity can be developed with
the help of dicrete subgroups of the group of all FLT \cite{Maskit}, with the related
parabolic, elliptic, hyperbolic and loxodromic sectors.

\begin{appendix}
\section{The Weyl limit-point limit-circle criterion for self-adjointness}
\setcounter{equation}{0}

The function $V$ is in the {\it limit-circle at $0$} if for some $\lambda$, and therefore
all $\lambda$ \cite{Coddington}, all solutions of the equation
\begin{equation}
\left[ -{d^{2}\over dx^{2}}+V(x) \right]\varphi(x)=\lambda \; \varphi(x)
\label{(A1)}
\end{equation}
are square-integrable at $0$ \cite{Coddington,Reed}.
\vskip 0.3cm
\noindent
If $V$ is not in the limit-circle case at $0$, it is said to be in the
{\it limit-point at $0$}.
\vskip 0.3cm
\noindent
According to the Weyl limit-point limit-circle criterion, if $V$ is a continuous 
real-valued function on the open interval $(0,\infty)$, the operator
\begin{equation}
P \equiv -{d^{2}\over dx^{2}}+V(x)
\label{(A2)}
\end{equation}
is essentially self-adjoint on $C_{0}^{\infty}(0,\infty)$ 
(i.e. it is closable and its closure is
self-adjoint therein) if and only if $V$ is in the limit-point case both at
$0$ and at $\infty$ \cite{Reed,Simon4,Bellino}. 
\end{appendix}

\section*{Acknowledgments}

The authors are grateful to the Dipartimento di Fisica ``Ettore Pancini'' of Federico II 
University for hospitality and support. G. Esposito is grateful to V. F. Bellino for
conversations \cite{Bellino}.

\end{document}